\documentclass[12pt]{iopart}

\usepackage{xcolor}  
\usepackage{mathrsfs}
\usepackage{amssymb}
\usepackage{pifont}
\usepackage{graphicx}
\newcommand{\cmark}{\ding{51}}

\usepackage{hyperref}
\usepackage{cite}
\def\code#1{\texttt{#1}}
\usepackage{amsfonts}
\usepackage[T1]{fontenc}
\usepackage[utf8]{inputenc}

\begin{document}

\title{Characterization of anomalous diffusion classical statistics powered by deep learning (CONDOR)}

\author{Alessia Gentili, Giorgio Volpe}

\address{Department of Chemistry, University College London, 20 Gordon Street, London WC1H 0AJ, UK}
\ead{g.volpe@ucl.ac.uk}
\vspace{10pt}

\begin{abstract}
Diffusion processes are important in several physical, chemical, biological and human phenomena. Examples include molecular encounters in reactions, cellular signalling, the foraging of animals, the spread of diseases, as well as trends in financial markets and climate records. Deviations from Brownian diffusion, known as anomalous diffusion, can often be observed in these processes, when the growth of the mean square displacement in time is not linear. An ever-increasing number of methods has thus appeared to characterize anomalous diffusion trajectories based on classical statistics or machine learning approaches. Yet, characterization of anomalous diffusion remains challenging to date as testified by the launch of the Anomalous Diffusion (AnDi) Challenge in March 2020 to assess and compare new and pre-existing methods on three different aspects of the problem: the inference of the anomalous diffusion exponent, the classification of the diffusion model, and the segmentation of trajectories. Here, we introduce a novel method (CONDOR) which combines feature engineering based on classical statistics with supervised deep learning to efficiently identify the underlying anomalous diffusion model with high accuracy and infer its exponent with a small mean absolute error in single 1D, 2D and 3D trajectories corrupted by localization noise. 
Finally, we extend our method to the segmentation of trajectories where the diffusion model and/or its anomalous exponent vary in time.
\end{abstract}

\vspace{2pc}

\section{Introduction}

Anomalous diffusion refers to diffusion phenomena characterized by a mean square displacement (MSD) that grows in time with an exponent $\alpha$ that is either smaller (subdiffusion) or greater (superdiffusion) than one (standard Brownian diffusion) \cite{ReviewMetzler}. Typically, this is represented by a nonlinear power-law scaling (${\rm MSD}(t) \sim t^\alpha$)  \cite{ReviewMetzler}. The growing interest in anomalous diffusion shown by the physics community lies in the fact that it extends the concept of Brownian diffusion and helps describe a broad range of phenomena of both natural and human origin \cite{AnomalousWings}. Examples include molecular encounters in reactions \cite{KineticsCell}, cellular signalling \cite{CrowdCells,StrangeMD}, the foraging of animals \cite{ForagingBook}, search strategies \cite{Volpe2017}, human travel patterns \cite{HumanTravel, HumanMobility}, the spread of epidemics and pandemics \cite{Epidemic, Pandemic}, as well as trends in financial markets \cite{Economics, StockPrices} and climate records \cite{Climate}. 

Various methods have been introduced over time to characterize anomalous diffusion in real data. Traditional methods are based on the direct analysis of trajectory statistics \cite{Toolbox} via the examination of features such as the mean square displacement \cite{MSD_bacteria, MSD_intracell, MSD_fitting, MSD_Brownian}, the velocity autocorrelation function \cite{VAF_bacteria} and the power spectral density \cite{PSD, PSD2} among others \cite{Toolbox, MME_method, MomentRatio_worm, FIMA, ARFIMA, Bayesian1, Bayesian2, Bayesian3, Bayesian4, Bayesian5, Bayesian6, pvar, StatisticalTesting}. More recently, the blossoming of machine learning approaches has expanded the toolbox of available methods for the characterization of anomalous diffusion with algorithms based on random forests \cite{RandomForests,RF_GB}, gradient boosting techniques \cite{RandomForests}, recurrent neural networks \cite{RNN} and convolutional neural networks \cite{CNN, DL}. 

Despite the significant progress made, the characterization of anomalous diffusion data still present non-negligible challenges, especially to maintain the approach insightful, easy to implement and parameter-free \cite{Andi}. First, the measurement of anomalous diffusion from experimental trajectories is often limited by the lack of significant statistics (i.e. single and short trajectories), limited localization precision, irregular sampling as well as imperceptible time fluctuations and changes in diffusion properties. Then, these limitations become even more pronounced when dealing with non-ergodic processes \cite{ReviewMetzler}, as the information of interest can be well hidden within individual trajectories. Finally, while machine learning approaches are powerful and parameter-free, more traditional statistics-based approaches typically offer deeper insight on the underlying diffusion process.

These shortcomings have resulted in the launch of the Anomalous Diffusion (AnDi) Challenge by a team of international scientists in March 2020 (\url{http://www.andi-challenge.org}) \cite{Andi}. The goal of the challenge was to push the methodological frontiers in the characterization of three different aspects of anomalous diffusion for 1D, 2D and 3D noise-corrupted trajectories \cite{Andi}: (1) the inference of the anomalous diffusion exponent $\alpha$; (2) the classification of the diffusion model; and (3) the segmentation of trajectories characterized by a change in anomalous diffusion exponent and/or model at an unspecified time instant.

Here, in response to the AnDi Challenge, we describe a novel method (CONDOR:~Classifier Of aNomalous DiffusiOn tRajectories) for the characterization of single anomalous diffusion trajectories based on the combination of classical statistics analysis tools and supervised deep learning. Specifically, our method applies a deep feed-forward neural network to cluster parameters extracted from the statistical features of individual trajectories. By combining advantages from both approaches (classical statistics analysis and machine learning), CONDOR is highly performant and parameter-free. It allows us to identify the underlying anomalous diffusion model with high accuracy ($\gtrsim 84\%$) and infer its exponent $\alpha$ with a small mean absolute error ($\lesssim 0.16$) in single 1D, 2D and 3D trajectories corrupted by localization noise. CONDOR was always among the top performant methods in the AnDi Challenge for both the inference of the $\alpha$-exponent and the model classification of 1D, 2D and 3D anomalous diffusion trajectories. In particular, CONDOR was the leading method for the inference task in 2D and 3D and for the classification task in 3D. 
Finally, in this article, we extend CONDOR to the segmentation of trajectories.

\begin{figure}
\centering
\includegraphics[width=1.\textwidth]{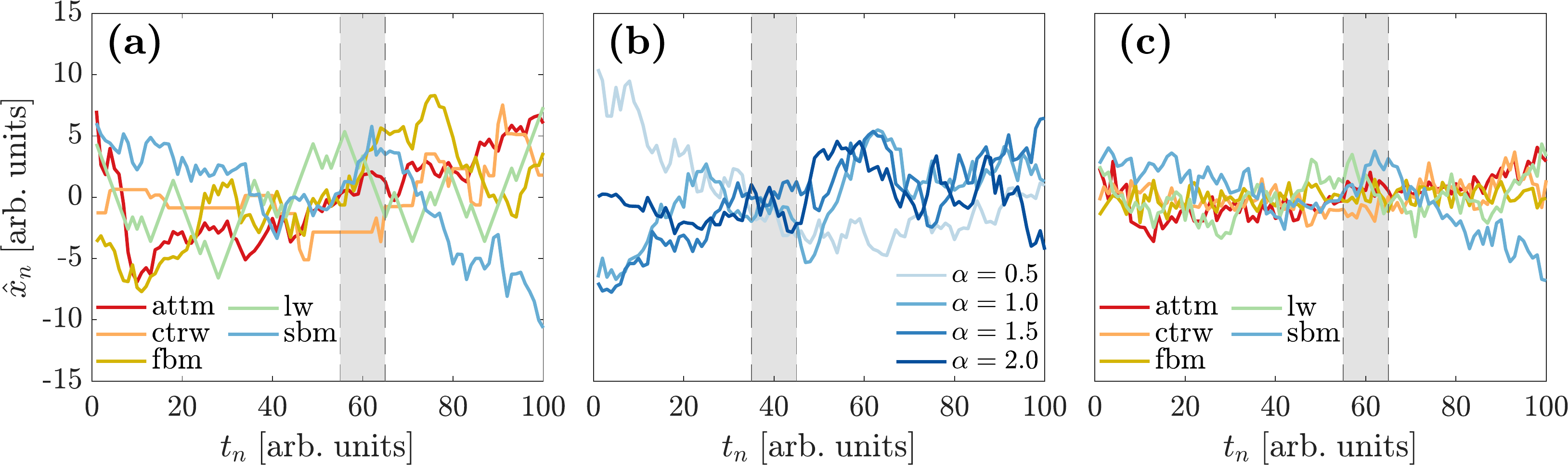}
\caption{{\bf Representative anomalous diffusion trajectories.} \textbf{(a-b)} Examples of single 1D anomalous diffusion trajectories in the absence of corrupting noise for \textbf{(a)} different models of standard diffusion (anomalous diffusion exponent $\alpha = 1$) and for \textbf{(b)} the same anomalous diffusion model (sbm) with different values of $\alpha$. The models considered here are the annealed transient time motion (attm), the continuous-time random walk (ctrw), the fractional Brownian motion (fbm), the L\'evy walk (lw) and the scaled Brownian motion (sbm). \textbf{(c)} Same trajectories as in \textbf{(a)} corrupted with Gaussian noise (of zero mean and variance $\sigma_n = 1$) associated to a finite localization precision, corresponding to a signal-to-noise ratio (SNR) of 1. For better comparison, each sequence of positions $x_n$ taken at times $t_n$ is represented after subtracting its average and after rescaling its displacements to their standard deviation to give the normalized sequence $\hat{x}_n$ \cite{RandomForests}. The grey-shaded areas highlight a small portion (10 data points) of each trajectory.}
\label{fig:fig1}
\end{figure}

\section{Methods}

CONDOR combines classical statistics analysis of single anomalous diffusion trajectories with supervised deep learning. The trajectories used in this work are generated numerically (with the \code{andi-dataset} Python package \cite{dataset}) from the 5 models considered in the AnDi Challenge: the annealed transient time motion (attm) \cite{ATTM}, the continuous-time random walk (ctrw) \cite{CTRW}, the fractional Brownian motion (fbm) \cite{FBM}, the L\'evy walk (lw) \cite{LW} and the scaled Brownian motion (sbm) \cite{SBM}). Briefly, AnDi Challenge datasets contain 1D, 2D or 3D trajectories from these 5 models sampled at regular time steps ($t_n = n\Delta t$, with $n$ a positive integer and $\Delta t$ a unitary time interval). The anomalous exponent $\alpha$ varies in the range $[0.05, 2]$ with steps of 0.05, as allowed by the respective model \cite{Andi}. The shortest trajectories have at least 10 samples per dimension ($T_{\rm min}$). Prior to the addition of the localization error, the trajectories have a unitary standard deviation $\sigma_s$ of the distribution of the displacements. The trajectories are then corrupted with additive Gaussian noise (with zero mean and variance $\sigma_n$) associated to a finite localization precision to better simulate experimental endeavours \cite{Andi, dataset}. The variance $\sigma_n$ of the Gaussian noise can assume values of 0.1, 0.5 or 1.0 along each direction of the trajectories. The final level of noise can be computed as the magnitude of a vector whose elements are the noise levels along each direction \cite{dataset}. The inverse of this value is then used to compute the signal-to-noise ratio (SNR). Figure 1 shows examples of such trajectories without noise (Figure~\ref{fig:fig1}a-b) and with the addition of corrupting noise (Figure~\ref{fig:fig1}c). By visual inspection, these examples show how differences between single trajectories are not always explicit, thus making the characterization of anomalous diffusion challenging. Already in the absence of noise, the same diffusion process (e.g. for $\alpha = 1$ in Figure~\ref{fig:fig1}a) can be described by different models and the same model (e.g. sbm in Figure~\ref{fig:fig1}b) can assume different values of the exponent $\alpha$. These occurrences are not always easy to set apart (e.g. attm, fbm and sbm in Figure~\ref{fig:fig1}a). The presence of corrupting noise (Figure~\ref{fig:fig1}c) and/or short trajectories (shaded areas in Figure~\ref{fig:fig1}a-c) add extra complexity.

\begin{figure}
\centering
\includegraphics[width=1\textwidth]{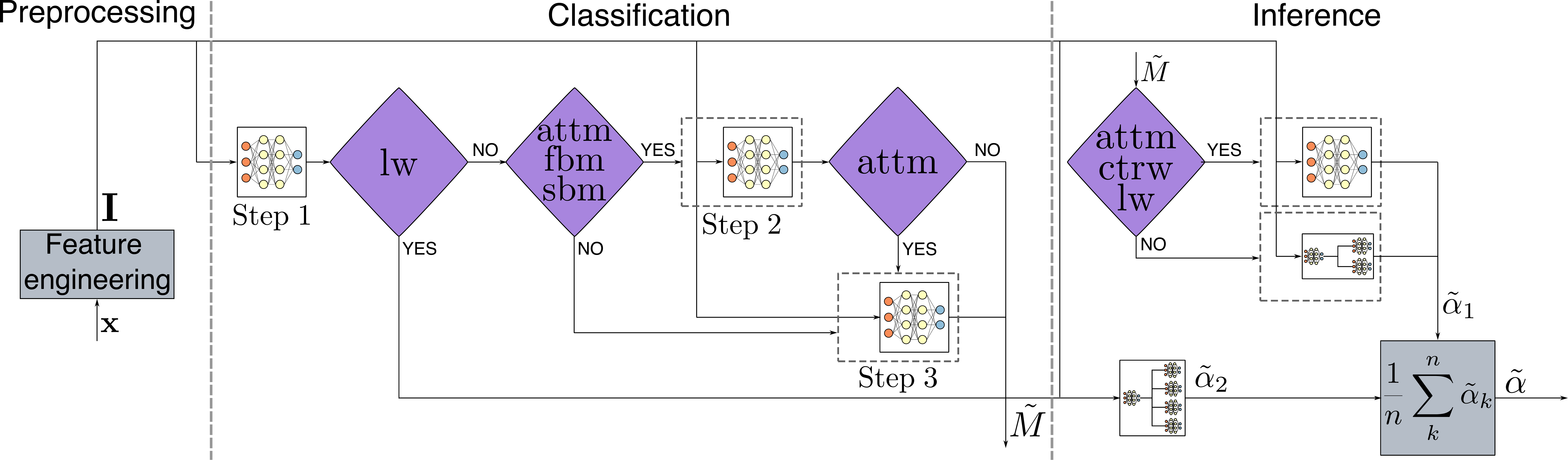}
\caption{\textbf{CONDOR workflow.}  CONDOR workflow is divided in three main parts: preprocessing (Section~\ref{sec:preprocess}), classification and inference (Section~\ref{sec:NetArch}). In the preprocessing step, the vector of positions $\bf{x}$ associated to a single trajectory is used to extract a vector $\bf{I}$ of $N = 1+92d$ inputs primarily based on classical statistics (with $d=1,2,3$ being the trajectory dimension). The vector $\bf{I}$ is the basic input for each neural network in the classification and inference algorithms. The output of the classification is determined by the output of up to three deep feed-forward neural networks and is the predicted model $\tilde{M}$ among five possibile categories: the annealed transient time motion (attm), the continuous-time random walk (ctrw), the fractional Brownian motion (fbm), the L\'evy walk (lw) and the scaled Brownian motion (sbm). Finally, the inference algorithm predicts the value of the exponent $\alpha \in [0.05,\, 2] $ basing its decision on $\bf{I}$ and the predicted $\tilde{M}$ for each trajectory. In particular, the predicted $\tilde{\alpha}$ is the arithmetic average of two distinct predictions: $\tilde{\alpha}_1$ and $\tilde{\alpha}_2$ where $\tilde{M}$ is used as selection element or additional input for the deep learning algorithm, respectively.}
\label{fig:fig2}
\end{figure}

In this section, we provide a detailed description of our method (CONDOR) to characterise such single anomalous diffusion trajectories. The fundamental intuition behind our approach is that different trajectories produced by the same diffusion model should share similar statistical features to set them apart from other models \cite{ReviewMetzler}. Nonetheless, in the presence of noise or with few data points, these similarities and differences do not need be explicit in single trajectories (Figure~\ref{fig:fig1}). Thus, CONDOR first manipulates single trajectories to extract this hidden statistical information and then employs the power of deep learning to analyse and cluster it. CONDOR workflow is shown in Figure~\ref{fig:fig2}. We will first discuss the feature engineering step used to preprocess the trajectories and extract a set of $N$ relevant features primarily based on classical statistics analysis (Section~\ref{sec:preprocess}). These features become the inputs for the deep-learning part of the characterization algorithm based on deep feed-forward neural networks (Figure \ref{fig:fig3}), which, in order, classifies the diffusion model and infers the anomalous diffusion exponent $\alpha$ (Figure~\ref{fig:fig2}). These two steps are performed by multiple neural networks, whose architecture and connection is explained in Section~\ref{sec:NetArch}. Section~\ref{sec:NetTrain} discusses the training of CONDOR neural networks. In Section~\ref{sec:NetSegm}, we explain how CONDOR outputs can be used to perform the segmentation of trajectories, where the model and/or its anomalous exponent change in time. Finally, in Section~\ref{sec:codes}, we provide a brief explanation of the code behind CONDOR \cite{code}.

\subsection{Feature engineering and CONDOR input definition}
\label{sec:preprocess}

Before the classification and inference steps (Figure~\ref{fig:fig2}), we analyzed each trajectory to extract $N = 1+92d$ inputs (with $d=1,2,3$ being the trajectory dimension) primarily based on classical statistics (Table \ref{table:inputs1}). The inputs used in this work were ultimately selected based on the final performance of the deep learning algorithm.

Independently of the dimensionality of the problem, the first input (${\rm ln} T_{\rm max}$) is always connected to the duration $T_{\rm max}$ of each trajectory to enable the deep learning algorithm to properly weight different durations. 

To extract the remaining $92d$ inputs for every trajectory in 1D, 2D and 3D, we analysed the vector of the positions along each direction $\textbf{x}^i \equiv (x^i_1, x^i_2, .... \, x^i_{\frac{T_{\rm max}}{\Delta t}}) $ (with $i = 1, ... \, d$) independently, after first subtracting its average and rescaling its displacements $w_j^i = x_{j+1}^i - x_j^i$ (with $j = 1, ... \, \frac{T_{\rm max}}{\Delta t}-1$) to their standard deviation to obtain the normalised displacements $\hat{w}_j^i$ \cite{RandomForests}. Practically, for each dimension, this step produces new time sequences $\hat{\textbf{x}}^i \equiv (\hat{x}^i_1, \hat{x}^i_2, .... \, \hat{x}^i_{\frac{T_{\rm max}}{\Delta t}})$ of positions reconstructed from the cumulative sum of the normalised displacements of the original trajectories and with $\langle \hat{\textbf{x}}^i \rangle = \textbf{0}$. Importantly, this step allows for a better comparison between trajectories of different dimensions and durations while keeping the performance of the deep learning algorithm optimal \cite{RandomForests}. 

The vectors of the normalised positions $\hat{\textbf{x}}^i$ and of the normalised displacements $\hat{\textbf{w}}^i \equiv (\hat{w}^i_1, \hat{w}^i_2, .... \, \hat{w}^i_{\frac{T_{\rm max}}{\Delta t}-1})$ are the starting point to calculate the statistical information (i.e. the deep learning inputs) associated to each trajectory. In particular, for each dimension, we calculated the mean ($\mu^i$), the median ($\mu_{1/2}^i$), the standard deviation ($\sigma^i$), the skewness ($\beta_1^i$), the kurtosis ($\beta_2^i$) and the approximate entropy (${\rm ApEn}^i$) associated to the following (Table \ref{table:inputs1}): the normalised displacements $\hat{\textbf{w}}^i$ and the vector of their absolute values $\hat{\textbf{W}}^i \equiv (|\hat{w}^i_1|, |\hat{w}^i_2|, .... \, |\hat{w}^i_{\frac{T_{\rm max}}{\Delta t}-1}|)$, the vector ${\bf \Delta}^i(m)$ of the absolute displacements $\Delta_j^i(m) = |\hat{x}_{j+m}^i - \hat{x}_j$| taken at time lags $\tau = m\Delta t$ (with $m = 2, 3,..., \frac{T_{\rm min}}{\Delta t}-1$), the vector $\textbf{r}^i$ of the ratios $r^i_j =  \frac{\hat{w}_{j+1}^i}{\hat{w}_j^i}$, the absolute values $|\mathcal{F}^i \{\hat{\textbf{w}}^i\}|$ of the normalized Fourier transform of $\hat{\textbf{w}}^i$, the vector $\textbf{P}_{\delta}^i$ of the time series of the normalised powers of  $\hat{\textbf{w}}^i$ segments selected with a non-overlapping moving window of width $\delta \Delta t $ (with $\delta = 3$), the vector $\mathbf{\Delta MSD}^i$ of the absolute values of the normalised variations $\Delta {\rm MSD}_j^i = |\frac{{\rm MSD}_{j+1}^i - {\rm MSD}_j^i}{j^2\Delta t^2}|$ of the time-averaged mean square displacement (MSD), and the absolute values ${\rm |\mathcal{W}}_u^i \{\textbf{$\hat{\textbf{w}}^i$}\}|$ of the wavelet transform of  $\hat{\textbf{w}}^i$ at the first two translations $u\Delta t$ with $u = 1, 2$.

To further improve the performance of the deep learning algorithm, the following extra four inputs were also added for each dimension to extract additional information about regular and irregular time patterns in the trajectories (correlations, power variations and abrupt changes): the integral of the normalized autocorrelation function (${\rm ACF}$) of $\hat{\textbf{w}}^i$ over a time window of width $\delta\Delta t$ to the right of the autocorrelation peak, the level of frequency asymmetry in the normalized power spectral density of $\hat{\textbf{w}}^i$ calculated as $(\frac{\Delta t}{T_{\rm max}})^2\sum_{k} {\rm |\mathcal{F}^\textit{i} \{\hat{\textbf{w}}}^i\}_k |^2 {\rm sgn}(k- \frac{F_{\rm N}}{2})$ with $F_{\rm N}$ being the Nyquist frequency, the level of time asymmetry in $\textbf{P}_{\delta}^i$ calculated as $\sum_{k} P_{\delta,k}^i {\rm sgn}(k- \left \lceil{\frac{T_{\rm max}}{2\delta}}\right \rceil)$, and the occurrence of abrupt changes in the variance of the elements of $\hat{\textbf{x}}^i $ \cite{Killick2012}.

\begin{table}
\caption{\label{table:inputs1} Statistical information extracted from each dimension of every trajectory as inputs for the deep learning algorithm. The approximate entropy is introduced as a way to quantify the amount of regularity and unpredictability of fluctuations over the considered time series. $m = 2, 3,..., \frac{T_{\rm min}}{\Delta t}-1$ and $u = 1, 2$. Note that ${\rm ApEn}^i$ is not defined for ${\bf \Delta}^i(m)$ when $m = \frac{T_{\rm min}}{\Delta t}-2$ and $m = \frac{T_{\rm min}}{\Delta t}-1$.}
\footnotesize\rm
\begin{tabular*}{\textwidth}{@{}l*{15}{@{\extracolsep{0pt plus12pt}}l}}
\br
 &  $\hat{\textbf{w}}^i$ &  $\hat{\textbf{W}}^i$ &  ${\bf \Delta}^i(m)$ &  ${\bf r}^i$ & $|\mathcal{F}^i \{\hat{\textbf{w}}^i\}|$ & $\textbf{P}_{\delta}^i$ & $\mathbf{\Delta MSD}^i$ & ${\rm |\mathcal{W}}_u^i \{\hat{\textbf{w}}^i\}|$ \\
\mr
$\mu^i$ & \cmark & \cmark & \cmark & & \cmark & \cmark & \cmark & \cmark \\
$\mu_{1/2}^i$ & \cmark & \cmark & \cmark & \cmark & \cmark & \cmark & \cmark & \cmark \\
$\sigma^i$ & & \cmark & \cmark & & \cmark & \cmark & \cmark & \cmark  \\
$\beta_1^i$ & \cmark & \cmark & \cmark & & \cmark & \cmark & \cmark & \cmark  \\
$\beta_2^i$ & \cmark & \cmark & \cmark & & \cmark & \cmark & \cmark & \cmark  \\
${\rm ApEn}^i$ & \cmark & \cmark & (\cmark) & & \cmark & \cmark & \cmark & \cmark  \\
\br
\end{tabular*}
\end{table}

\subsection{CONDOR architecture}\label{sec:NetArch}

\begin{figure}
\centering
\includegraphics[width=1\textwidth]{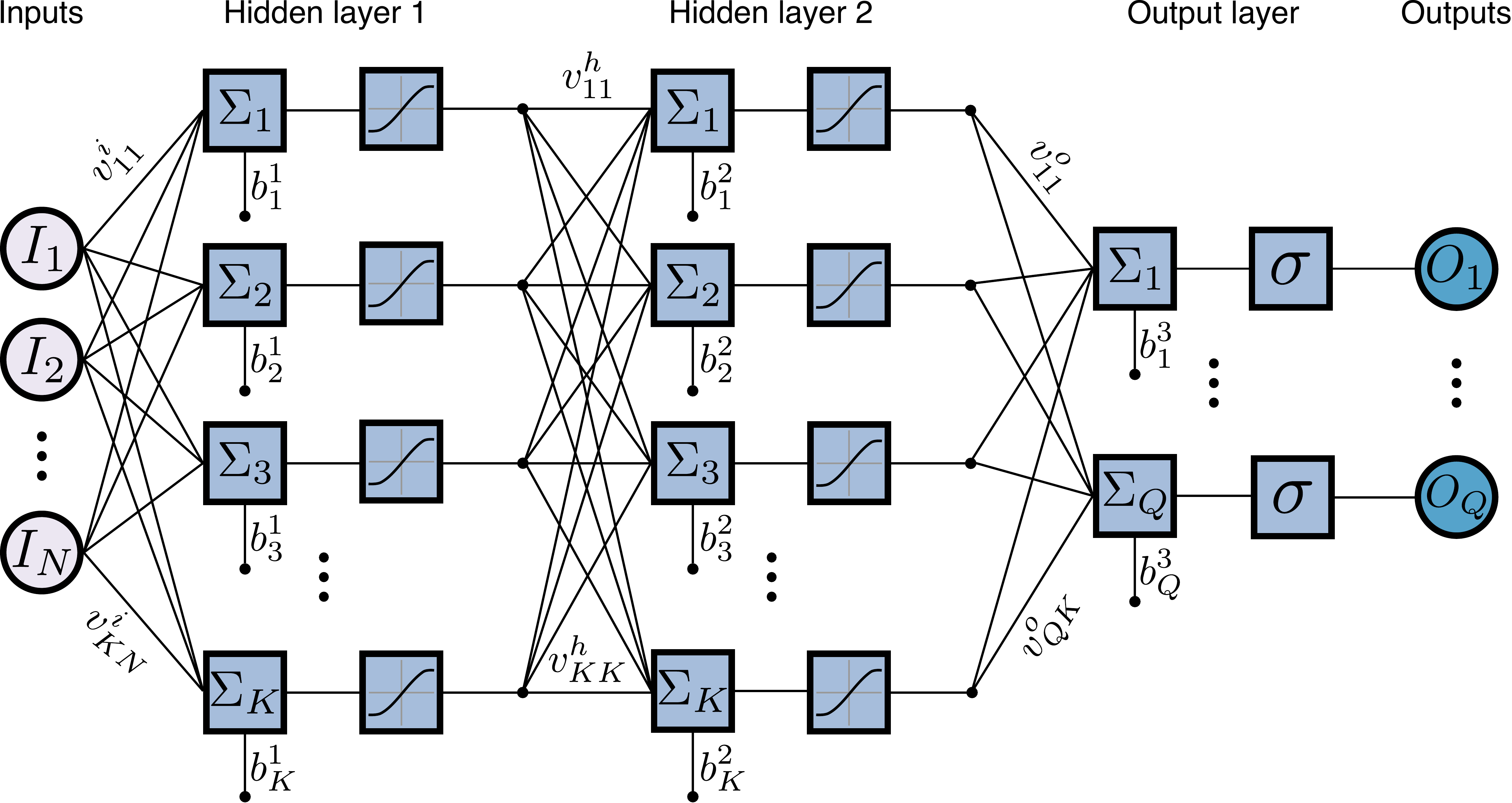}
\caption{{\bf Basic neural network architecture in CONDOR.} Architecture of the deep three-layer feed-forward neural network chosen for both classification and inference tasks. Inputs $I_n$ (with $n = 1,2,...N$) are processed by two hidden layers and an output layer to yield outputs $O_q$ (with $q = 1,2,...Q$). The hidden layers have $K = 20$ neurons each with a sigmoidal activation function ($\int$). The output layer has $Q$ neurons with a softmax activation function ($\sigma$). Each neuron has a summer ($\mathsf{\Sigma}$) that gathers its weighted inputs and bias to return a scalar output. For each neuron, weights $v$ and bias $b$ are optimised during the training of the network.}
\label{fig:fig3}
\end{figure}

In the model classification task (Figure~\ref{fig:fig2}), the $N$ inputs calculated in the previous step are first fed to a deep three-layer feed-forward neural network (architecture in Figure \ref{fig:fig3}) with two sigmoid hidden layers of twenty neurones each and one softmax output layer of five neurones to predict the model $\tilde{M}$ of anomalous diffusion among the five possible categories (attm, ctrw, fbm, lw and sbm). We settled for this specific architecture with two dense layers because it is deep enough to efficiently classify noisy vectors arbitrarily while still avoiding that its learning rate is slowed down by an excessively complex architecture. The output of the previous neural network is then refined by two additional networks with similar architectures (but with three output neurones and two output neurones, respectively) to improve the classification among trajectories respectively classified as attm, fbm and sbm and as attm and ctrw (Figure~\ref{fig:fig2}), as these models can be more easily confused. Indeed, in the attm model, the diffusion coefficient can vary in space or time (in the AnDi challenge only time variations were considered \cite{ATTM, Andi}). This feature can be easily confused with features of other processes: namely, the time-dependent diffusivity of the sbm \cite{SBM}, the time irregularities between consecutive steps in the ctrw \cite{CTRW} and the correlations among the increments in the fbm \cite{FBM}.

Once the model is identified, the next step is to infer the anomalous diffusion exponent $\alpha$. The estimated value $\tilde{\alpha}$ is obtained as the arithmetic average of the outputs of two different approaches based on deep learning (Figure~\ref{fig:fig2}), as the two approaches lead to predicting an underestimated $\alpha$ value ($\tilde{\alpha}_1$) and an overestimated  $\alpha$ value ($\tilde{\alpha}_2$), respectively. In the first approach (Figure~\ref{fig:fig2}), the output $\tilde{M}$ of the classification is used to cluster the trajectories into five categories based on the identified model. Each category is analysed independently by a different set of neural networks (architectures as in Figure \ref{fig:fig3}). The neural networks for attm and ctrw have five output categories predicting $\tilde{\alpha}_1$ in the range 0.05 to 1 \cite{Andi, dataset}. The neural network for lw has five output categories predicting $\tilde{\alpha}_1$ in the range 1 to 2 \cite{Andi, dataset}. Finally, the prediction for fbm and sbm is based on a 1x2 directed rooted tree of neural networks with the root network having two equally spaced output categories predicting $\tilde{\alpha}_1$ in the range 0.05 to 2; each category is then refined by a second neural network with five equally spaced output categories in the corresponding subrange identified by the root network. In the second approach (Figure~\ref{fig:fig2}), the output $\tilde{M}$ of the model classification is used as an additional input together with the list of inputs identified in Section~\ref{sec:preprocess}. These new set of inputs feed a 1x4 directed rooted tree of neural networks, each with the same overall architecture as in Figure \ref{fig:fig3}. The root neural network has four equally spaced output categories predicting $\tilde{\alpha}_2$ in the range 0.05 to 2; each of this category is then refined by a second neural network with five equally spaced output categories in the corresponding $\alpha$ subrange identified by the root network. 

\begin{figure}
\centering
 \includegraphics[width=0.8\textwidth]{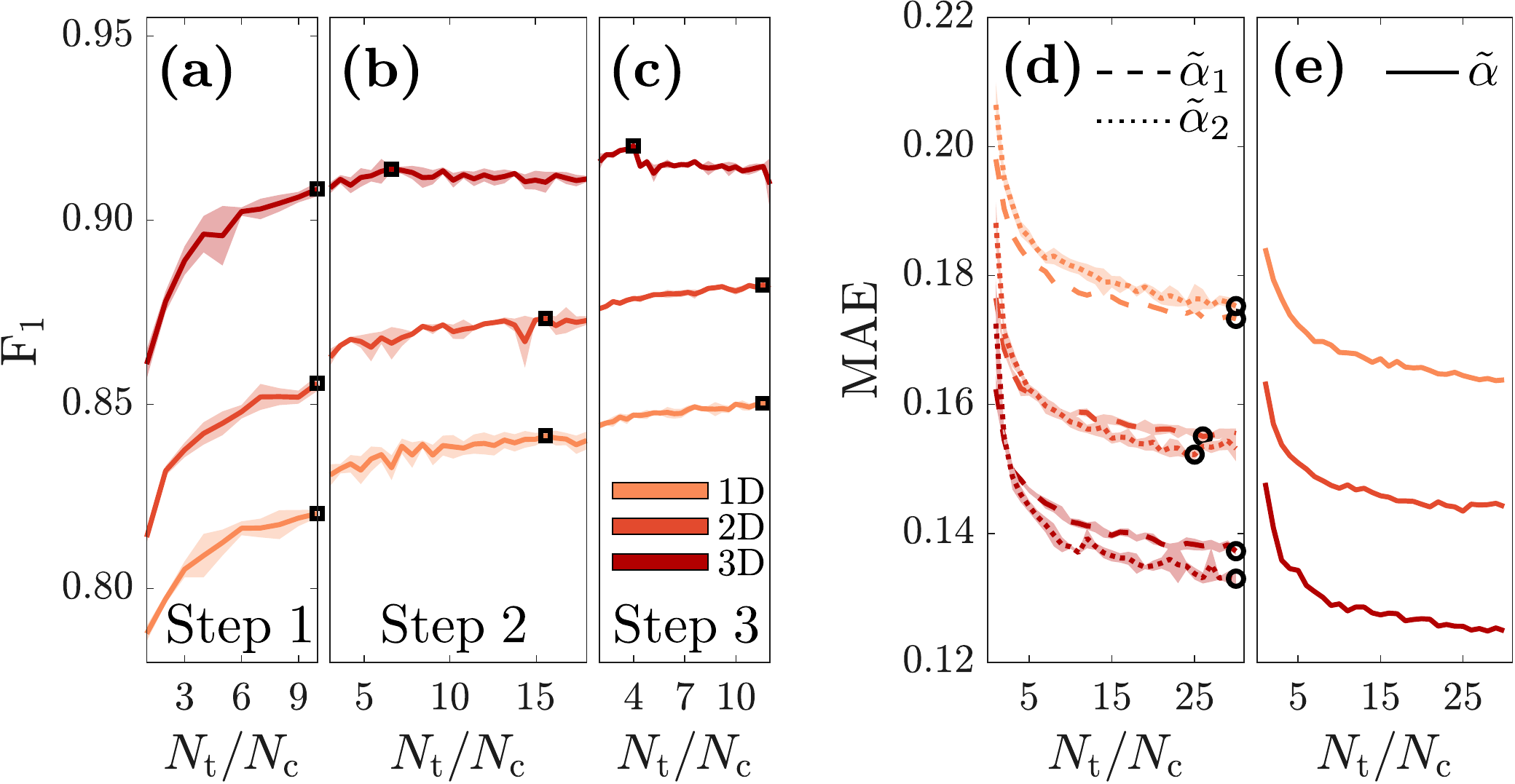}
\caption{\textbf{Evolution of CONDOR performance with the size of the training dataset.} \textbf{(a-c)} $\mathrm{F_1}$ score calculated on the total 1D, 2D and 3D training datasets of 300k trajectories as a function of the number of trajectories ($N_{\mathrm{t}}$) employed to train the three consecutive networks of the classification task (Figure~\ref{fig:fig2}). $N_{\mathrm{t}}$ is normalized to the size of a single AnDi challenge dataset ($N_{\mathrm{c}} =$ 10k trajectories).  The $\mathrm{F_1}$ score is given at the output of \textbf{(a)} the first network, \textbf{(b)} the second network and \textbf{(c)} the third network used to refine the model predictions in the classification task (Figure~\ref{fig:fig2}). At each of these steps, only the best performing network (black squares) is carried forward to the next step. \textbf{(d-e)} MAE calculated on the total 1D, 2D and 3D training datasets of 300k trajectories each as a function of $N_{\mathrm{t}}$/$N_{\mathrm{c}}$ for \textbf{(d)} 
$\tilde{\alpha}_1$ (dashed lines) and $\tilde{\alpha}_2$ (dotted lines) and \textbf{(e)} for $\tilde{\alpha}$ (solid lines) as defined in Figure~\ref{fig:fig2}. For each $N_{\mathrm{t}}$/$N_{\mathrm{c}}$, the arithmetic average $\tilde{\alpha}$ is calculated using the best performing networks used to infer $\tilde{\alpha}_1$ and $\tilde{\alpha}_2$ with the same number of training trajectories. The networks with the lowest MAE for $\tilde{\alpha}_1$ and $\tilde{\alpha}_2$ are also shown in \textbf{(d)} (black circles). The shaded areas correspond to one standard deviation around the mean values (calculated from the performance of 5 distinct training attempts).}
\label{fig:fig4}
\end{figure}

\subsection{Training of CONDOR}\label{sec:NetTrain}

After defining their architecture, we trained each network of CONDOR independently on a set of trajectories for which the ground-truth values of the models and of $\alpha$ was known. Each training was performed using the scaled conjugate gradient backpropagation (scg) algorithm (for speed) on mixed datasets with up to three hundred thousand trajectories from different models and values of $\alpha$, with different durations and levels of corrupting noise \cite{dataset}. Datasets were randomly split among training (70\%), validation (15\%) and test (15\%) trajectories. Overall, training with scg is very efficient on a standard desktop computer (processor Intel(R) Core(TM) i7-4790 CPU @ 3.60 GHz), being the size of the training dataset the main limiting factor. Training the three consecutive networks used for classification (Figure \ref{fig:fig2}) took at most 1 hour and a half, while training the fourteen networks for the inference task took at most 3 hours. Training time can be reduced by up to 2 orders of magnitude employing a GPU. After training, CONDOR is very computationally efficient on new data. For example, CONDOR can classify $\approx 80{\rm k}$ trajectories/s on a standard desktop computer.

Figure \ref{fig:fig4} shows the performance of the classification task and of the inference task with the size of the training dataset. We evaluated the performance of the classification with the $\mathrm{F_1}$ score:
\begin{equation*} \label{F1}
\mathrm{F_1} = \frac{TP}{TP+\frac{1}{2}(FP+FN)}
\end{equation*}
where $TP$, $FP$ and $FN$ are the true positive, false positive and false negative rates, respectively. In particular, we computed the \textit{micro} average of the $\mathrm{F_1}$ score with the Scikit-learn Python’s library \cite{scikit-learn}.
$\rm{F_1}$ can vary between 0 (no trajectory classified correctly) and 1 (all trajectories classified correctly). To evaluate the performance of the inference of the $\alpha$-exponent, we used the mean absolute error ($\rm{MAE}$) instead: 
\begin{equation*} \label{MAE}
\mathrm{MAE} = \frac{1}{n}\sum_{i=1}^n|\tilde{\alpha_i}-\alpha_i| 
\end{equation*}
where $n$ is the number of trajectories in the dataset, and $\tilde{\alpha}_i$ and $\alpha_i$ are the predicted and the ground truth values of the anomalous diffusion coefficient of the $i$-th trajectory, respectively. MAE has a lower limit of 0, meaning that the coefficients of all the $n$ trajectories have been predicted correctly.

Figure \ref{fig:fig4}a-c shows the $\rm{F_1}$ score measured on the training dataset at the output of the three consecutive networks used for classification (Figure \ref{fig:fig2}) for the 1D, 2D and 3D cases. For each step, we increased the number of trajectories used for training progressively until the $\rm{F_1}$ score reached near saturation. Past this point any gain in $\rm{F_1}$ score is relatively small and by far outbalanced by the increased computational cost associated to longer training times. The performance improves with the dimension $d$ of the trajectories and, interestingly, at each step, saturation of the $\rm{F_1}$ score for higher $d$ appears to take place with smaller training datasets, as each trajectory contains intrinsically $d$ times more information. Only the best performing network at each step is selected (black squares in Figure \ref{fig:fig2}a-c) and carried forward to the next step. Independently of the dimensionality of the problem, the first network produces the highest increase of $\rm{F_1}$ score with the size of the training dataset (Figure \ref{fig:fig4}a): for example, going from 10k to 100k trajectories, the $\rm{F_1}$ score increases approximately by 4\% in 1D, 5\% in 2D and 6\% in 3D. The combined contribution of the following two networks to the overall $\rm{F_1}$ score is less pronounced, especially for higher dimensions (Figure \ref{fig:fig4}b-c); nonetheless they play a key role in refining the classification among attm, fbm and sbm (Figure \ref{fig:fig4}b) or among attm and ctrw (Figure \ref{fig:fig4}c), thus boosting the final $\rm{F_1}$ score approximately by 4\% in 1D, 3\% in 2D and 1\% in 3D.

Similarly, Figure \ref{fig:fig4}d-e shows the ${\rm MAE}$ associated to the inference of the $\alpha$-exponent according to $\tilde{\alpha}_1$, $\tilde{\alpha}_2$ and their arithmetic average $\tilde{\alpha}$ (as defined in Figure \ref{fig:fig2}) for the 1D, 2D and 3D cases. As for ${\rm F}_1$, we report the $\rm{MAE}$ of each method as a function of the size of the training dataset progressively until it reaches near saturation. In general, the MAE lowers with the number of trajectories used for training and with their dimension $d$. As noted earlier, independently of the dimensionality of the problem, the arithmetic average $\tilde{\alpha}$ provides a better estimate of the $\alpha$ exponent with respect to both $\tilde{\alpha}_1$ and $\tilde{\alpha}_2$, leading to an overall MAE at saturation that is better than those obtained with the individual methods by at least 0.01 in all the dimensions.

\subsection{Segmentation with CONDOR}\label{sec:NetSegm}

The output of the classification and the inference can also be used to identify possible change points in a trajectory. In practice, a trajectory of length $T_{\rm max}$ is divided in $T_{\rm max} - B +1$ smaller segments using an overlapping moving window of width $B$ (with $B <<  T_{\rm max}$). The width of the moving window is a trade-off between the possibility of identifying change points close to either end of a trajectory (which increases for smaller $B$) and the accuracy in predicting the correct time of the change point (which increase with $B$ due to smaller errors in predicting the model and the $\alpha$ exponent associated to each segment). Each of these segments is then fed to CONDOR for classification and inference, thus generating two time series of $T_{\rm max} - B +1$ values of the predicted model and $\alpha$ exponent. The occurrence of abrupt changes (up to two) in the root-mean-square level of these time series is used to identify the point of change of the model and/or the $\alpha$ exponent \cite{Killick2012}. If two abrupt changes are identified, the overall change point is calculated to first approximation as the average between these two time instances.

To evaluate CONDOR's performance in identifying the presence of a change point, we calculated the root mean square error of the change point localization as in the AnDi Challenge \cite{Andi}:
\begin{equation*}
\mathrm{RMSE} = \frac{1}{n}\sqrt{\sum_{i=1}^n ( \tilde{t}_i - t_i)^2}
\end{equation*}
where $\tilde{t}_i$ is the prediction and ${t}_i$ the ground truth value of the change point, respectively, of a single trajectory in a dataset of $n$ trajectories. Similar to the MAE, the RMSE has a lower limit of 0 when all the change points of the $n$ trajectories are correctly identified.

\subsection{CONDOR codes}\label{sec:codes}

We have implemented all neural networks using the patternnet function in the MATLAB$^{\rm TM}$-based Deep Learning Toolbox$^{\rm TM}$. We provide the MATLAB$^{\rm TM}$ implementations of the key functionalities of CONDOR \cite{code}. Nonetheless, our approach is independent of the deep-learning framework used for its implementation. Moreover, it is also possible to exchange neural network models  between the Deep Learning Toolbox$^{\rm TM}$ and other platforms, such as those based on TensorFlow$^{\rm TM}$. 

\begin{figure}
\centering
\includegraphics[width=0.8\textwidth]{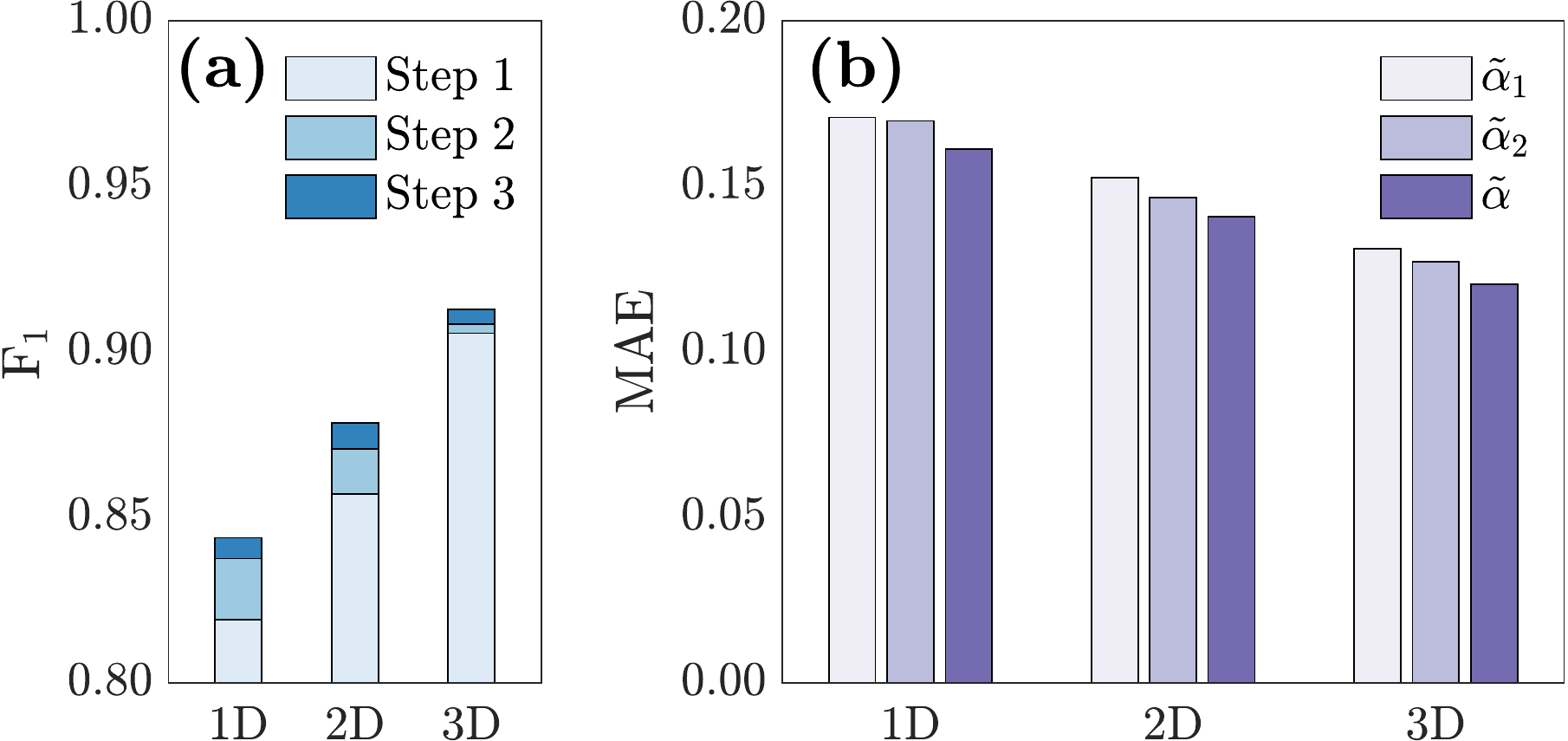}
\caption{\textbf{CONDOR performance on test datasets.} \textbf{(a)} $\mathrm{F_1}$ and \textbf{(b)} $\mathrm{MAE}$ obtained using CONDOR on the 1D, 2D and 3D datasets used in the classification and inference tasks of the AnDi Challenge (10k trajectories each). In \textbf{(a)}, the stacked bar charts shows the contribution of the three consecutive networks used for classification (Figure~\ref{fig:fig2}) to the total $\mathrm{F_1}$ score. In \textbf{(b)}, the grouped bar charts shows the MAE associated to the inference of the $\alpha$-exponent according to $\tilde{\alpha}_1$, $\tilde{\alpha}_2$ and their arithmetic average $\tilde{\alpha}$, which provides the best estimate of the $\alpha$ exponent (Figure \ref{fig:fig2}).} 
\label{fig:fig5}
\end{figure}

\section{Results}

\subsection{CONDOR overall performance on the datasets of the AnDi Challenge}

The best performing networks identified during the training step (black symbols in Figure \ref{fig:fig4}) were carried forward and used in the prediction step, where we classified the anomalous diffusion model and inferred the $\alpha$ exponent of the same datasets used in the AnDi Challenge (with 10k trajectories each) for testing purposes \cite{Andi}. Figure \ref{fig:fig5}a shows the overall $\rm{F_1}$ score obtained on these test databases: $\mathrm{F_1} = 0.844$, $\mathrm{F_1} = 0.879$ and $\mathrm{F_1} = 0.913$ for the 1D, 2D and 3D cases, respectively. Figure \ref{fig:fig5}b shows the overall $\rm{MAE}$ instead: $\mathrm{MAE} = 0.161$, $\mathrm{MAE} = 0.141$ and $\mathrm{MAE} = 0.121$ for the 1D, 2D and 3D cases, respectively. These very respectable values enabled CONDOR to be one of the top performant methods in the AnDi Challenge (\url{http://www.andi-challenge.org}) \cite{Andi}.

\subsection{Analysis of CONDOR performance in the model classification task}

\begin{figure}
\centering
\includegraphics[width=1\textwidth]{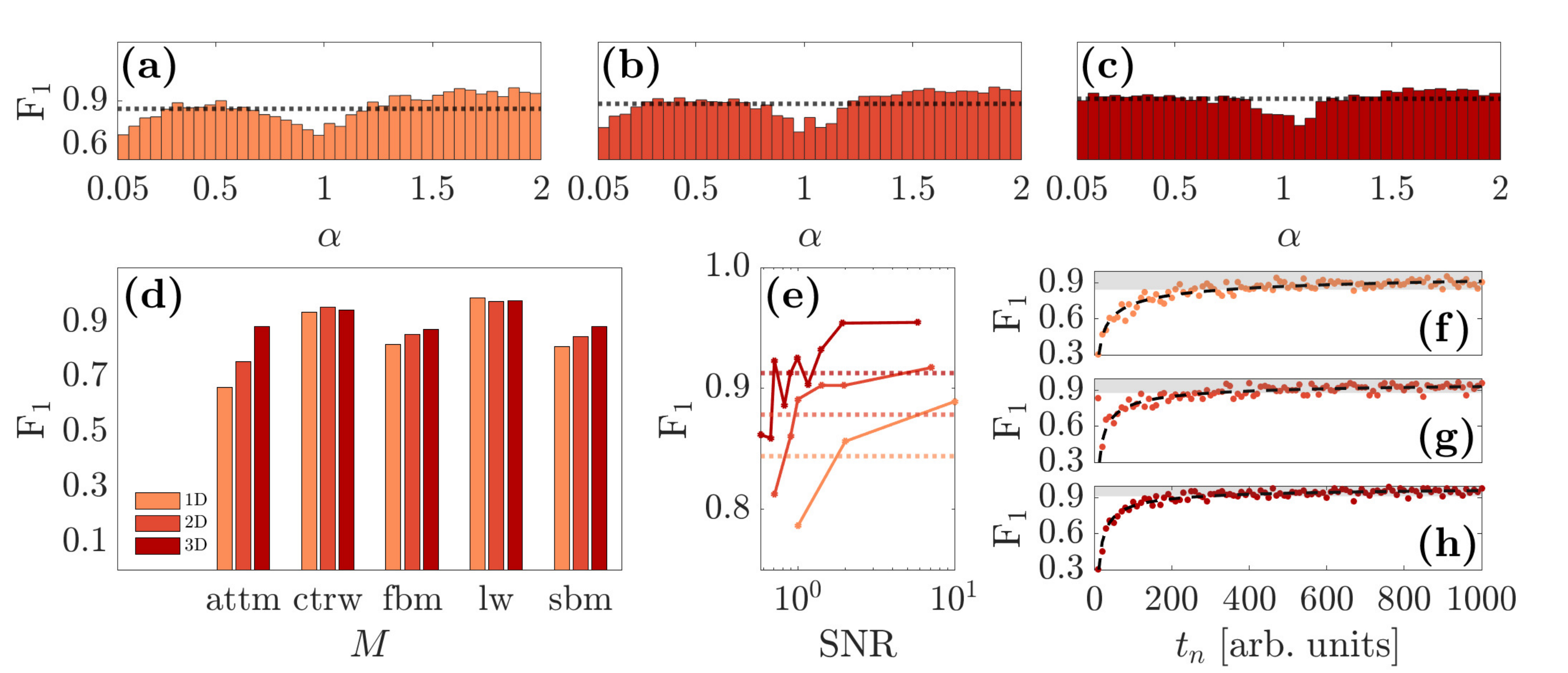}
\caption{\textbf{Detailed analysis of CONDOR performance for the classification task.} CONDOR classification performance ($\rm{F}_1$ score) on the 1D, 2D and 3D datasets of the AnDi Challenge used for the classification task (as in Figure \ref{fig:fig5}a) \textbf{(a-c)} by value of $\alpha$ for the \textbf{(a)} 1D, \textbf{(b)} 2D and \textbf{(c)} 3D case, \textbf{(d)} by model, \textbf{(e)} by signal-to-noise ratio (SNR), and \textbf{(f-h)} by trajectory length for the \textbf{(f)} 1D, \textbf{(g)} 2D and \textbf{(h)} 3D case, respectively. In \textbf{(a-c)} each bar corresponds to an increment in $\alpha$ of 0.05. In \textbf{(a-c)} and \textbf{(e)} the dotted lines represent the overall $\rm F_1$ as in Figure~\ref{fig:fig5}a. In \textbf{(f-h)}, the trends are fitted to a power law ($y = c-ax^{-b}$, with $a,b,c > 0$) for ease of visualization. The shaded areas highlight the values of the $\mathrm{F_1}$ score above the average value (Figure \ref{fig:fig5}a).}
\label{fig:fig6}
\end{figure}

Figure \ref{fig:fig6} provides further detail on CONDOR performance in the classification task by value of $\alpha$ (Figure \ref{fig:fig6}a-c), by model (Figure \ref{fig:fig6}d), by strength of the localization noise (Figure \ref{fig:fig6}e) and by trajectory length (Figure \ref{fig:fig6}f-h). These data largely support CONDOR robustness of performance across the parameter space.

CONDOR is able to predict the correct model relatively consistently across the values of $\alpha$, with the exception of $\alpha \simeq 0.05$ in 1D and 2D and of $\alpha \simeq 1$ for all dimensions (Figure \ref{fig:fig6}a-c). This is not surprising. In fact, for $\alpha \simeq 0.05$ all trajectories are close to stationary and show a very low degree of motion, which could be dominated by the localization noise. Therefore, these trajectories are harder to classify, particularly in lower dimensions where a smaller amount of information is intrinsically available. Similarly, for $\alpha \simeq 1$ all models converge to Brownian motion, making it much more challenging to distinguish among them. Interestingly, in 1D and 2D, CONDOR shows an above-than-average performance in classifying superdiffusive trajectories, thus compensating what lost in the classification of subdiffusive trajectories.  

Consistently across the dimensions (1D, 2D and 3D), CONDOR recognises two models (ctrw and lw) extremely well with ${\rm F_1 \geq 0.94}$ for ctrw and ${\rm F_1 \geq 0.98}$ for lw (Figure \ref{fig:fig6}d). This is understandable as both models show typical features that stand them apart from the remaining models and are relatively robust to the presence of localization noise (Figure \ref{fig:fig1}): i.e. long ballistic steps in lw trajectories and long waiting times in ctrw trajectories \cite{ReviewMetzler}. For the remaining three models (attm, fbm and sbm), CONDOR performs progressively better moving from 1D to 3D. As noticed earlier (Figure \ref{fig:fig1}), the distinctive features among these models are more subtle and can be better evaluated for higher dimensions as they are intrinsically richer in information content. While CONDOR performs equally well with fbm and sbm (${\rm F_1 \approx 0.81-0.82}$ in 1D, ${\rm F_1 \approx 0.85-0.86}$ in 2D and ${\rm F_1 \approx 0.87-0.89}$ in 3D), it struggles more with attm as this model can be more easily confused with others (Figure \ref{fig:fig1}). Nonetheless, CONDOR performance skyrockets with the dimension of the attm trajectories from ${\rm F_1 \approx 0.66}$ in 1D to ${\rm F_1 \approx 0.88}$ in 3D. 

The performance of CONDOR with different values of SNR gives the expected results (Figure \ref{fig:fig6}e). The ${\rm F_1}$ score increases for higher values of SNR, reaching ${\rm F_1}$ scores of 0.89, 0.92 and 0.95 for 1D, 2D and 3D, respectively (i.e. well above the overall ${\rm F_1}$ score in Figure \ref{fig:fig5}a). However, even when the noise overcomes the signal (SNR $\leq 1$), the lost in the classification performance is less than 10\% of these high values for the level of noise considered here.

Finally, Figure~\ref{fig:fig6}f-h shows CONDOR classification performance with the duration of the trajectory. Due to the richer information content, classification is increasingly better for longer trajectories and for higher dimensions. The trends in Figure~\ref{fig:fig6}f-h are well fitted by power law functions (dashed lines), which show how, above a certain threshold in duration, CONDOR performs above the average ${\rm F}_1$ scores (delimited by the shaded areas), reaching ${\rm F}_1$ values of up to 0.96 in 1D, 0.97 in 2D and 0.99 in 3D. Moreover, this threshold decreases when increasing the dimension, from $\approx 340$ time points in 1D to $\approx 280$ in 3D. Note that, by directly training CONDOR neural networks for classification only with short trajectories ($\leq$ 50 time steps), its performance in classifying them can be boosted. This possibility is particularly useful for the classification of experimental trajectories that often are short or are interrupted by missing data points \cite{RNN, SingleKinetics}. For example, we were able to improve the $\rm F_1$ score on short trajectories ($\leq$ 30 time steps) in Figure~\ref{fig:fig6}f-h by $\approx 15\%$ on average. 

\begin{figure}
\centering
\includegraphics[width=1.\textwidth]{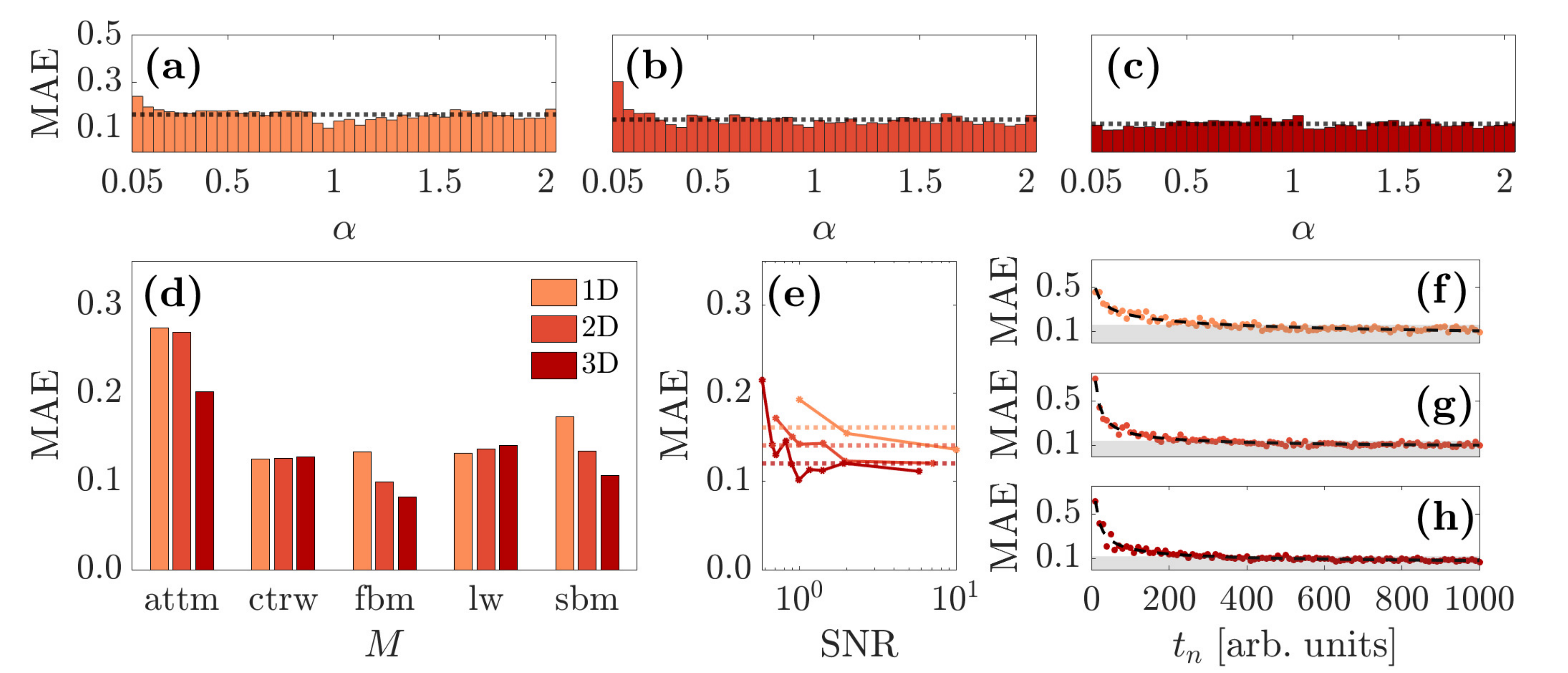}
\caption{\textbf{Detailed analysis of CONDOR performance for the $\alpha$-exponent inference task.} CONDOR $\alpha$-exponent inference performance ($\rm{MAE}$) on the 1D, 2D and 3D datasets of the AnDi Challenge used for the inference task (as in Figure \ref{fig:fig5}b) \textbf{(a-c)} by value of $\alpha$ for the \textbf{(a)} 1D, \textbf{(b)} 2D and \textbf{(c)} 3D case, \textbf{(d)} by model, \textbf{(e)} by signal-to-noise ratio (SNR), and \textbf{(f-h)} by trajectory length for the \textbf{(f)} 1D, \textbf{(g)} 2D and \textbf{(h)} 3D case, respectively. In \textbf{(a-c)} each bar corresponds to an increment in $\alpha$ of 0.05. In \textbf{(a-c)} and \textbf{(e)} the dotted lines represent the overall MAE as in Figure \ref{fig:fig5}b. In \textbf{(f-h)}, the trends are fitted to a power law ($y = ax^{-b}+ c$, with $a,b,c > 0$) for ease of visualization. The dashed areas highlight the values of the MAE below the average value (Figure \ref{fig:fig5}b).} 
\label{fig:fig7}
\end{figure}

\subsection{Analysis of CONDOR performance in the $\alpha$-exponent inference task}

Figure \ref{fig:fig7} provides further detail on CONDOR performance in the $\alpha$-exponent interference task by value of $\alpha$ (Figure \ref{fig:fig7}a-c), by model (Figure \ref{fig:fig7}d), by strength of the localization noise (Figure \ref{fig:fig7}e) and by trajectory length (Figure \ref{fig:fig5}f-h). These data largely confirm the observations on CONDOR performance for the classification task, thus confirming its robustness across the parameter space.

In Figure \ref{fig:fig7}a-c, we can see how CONDOR performs in a relatively consistent way across the $\alpha$ values, with the exception of $\alpha = 0.05$ in the 1D and 2D cases ($\rm MAE = 0.239 $ in 1D and $\rm MAE = 0.302$ in 2D). This is primarily due to a relatively large proportion of 1D and 2D attm trajectories with $\alpha = 0.05$ being assigned to other $\alpha$ values: for example, for these trajectories, $\approx10\%$ of the predicted values of $\alpha$ is greater than 1 (against only the $0.03\%$ of the 3D attm trajectories with $\alpha = 0.05$). As in the classification case, being these subdiffusive trajectories almost stationary, the inference is largely influenced by the localization noise.

As already noted for the model classification (Figure \ref{fig:fig6}a), CONDOR can also infer the $\alpha$-exponent very well and consistently across the dimensions for two models (ctrw and lw) with $\rm MAE \leq 0.128$ for ctrw and $\rm MAE \leq 0.141$ for lw (Figure \ref{fig:fig7}d). For the three remaining models (attm, fbm and sbm), CONDOR again performs progressively better moving from 1D to 3D. The best results across the dimensions are obtained for the fbm ($\rm MAE \approx 0.133$ in 1D, $\rm MAE\approx 0.100$ in 2D and $\rm MAE \approx 0.083$ in 3D), followed by sbm (${\rm MAE \approx 0.173}$ in 1D, ${\rm MAE\approx 0.134}$ in 2D and ${\rm MAE \approx 0.107}$ in 3D). Once again, as in the classification task, CONDOR struggles more with attm (${\rm MAE \approx 0.274}$ in 1D, ${\rm MAE\approx 0.269}$ in 2D and ${\rm MAE \approx 0.201}$ in 3D).

Figure \ref{fig:fig7}e confirms what has been discussed for the classification regarding CONDOR performance with different SNR values. For high SNRs, the MAE values for each dimension are well below the average MAE (Figure \ref{fig:fig5}b): $\rm MAE = 0.136$ in 1D, $\rm MAE = 0.121$ in 2D and $\rm MAE = 0.112$ in 3D. As expected, the MAE increases with higher levels of noise in all dimensions, but it is still below $\approx 0.2$ for the values of noise considered here.

Finally, Figure~\ref{fig:fig7}f-h shows CONDOR performance in inferring the $\alpha$ exponent with the duration of the trajectory. Similar to the classification task, CONDOR performance improves for higher dimensions and for longer trajectories as the MAE becomes increasingly lower in agreement with prior observations \cite{RNN, RandomForests}. The trends in Figure~\ref{fig:fig7}f-h are well fitted by power law functions (dashed lines), which show how, above a certain threshold in duration, CONDOR performance is significantly below the average MAE (delimited by the shaded areas), reaching values as small as 0.083 in 1D, 0.085 in 2D and 0.065 in 3D. As for the classification task, this threshold decreases when increasing the dimension, from $350$ time points in 1D to $300$ in 3D. Also for the inference task, it is possible to improve the prediction of the value of $\alpha$ for short trajectories ($\leq$ 40 time steps) by training CONDOR neural networks only using short trajectories ($\leq$ 50 time steps). Differently from the classification task, training CONDOR on short trajectories does not significantly reduce the error on the inference of the $\alpha$ exponent.

\section{Expanding CONDOR to the segmentation of trajectories}

\begin{figure}
\centering
\includegraphics[width=1.\textwidth]{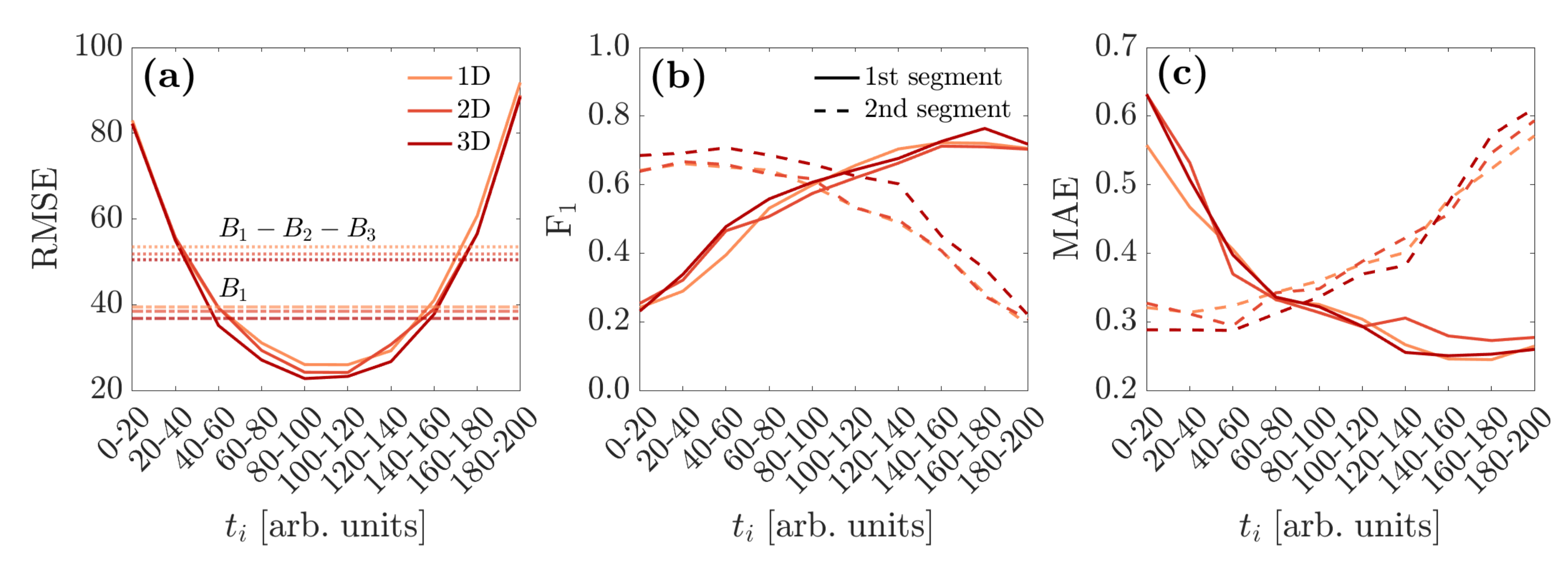}
\caption{\textbf{Detailed analysis of CONDOR performance for the segmentation task.} CONDOR segmentation results on the 1D, 2D and 3D datasets of the AnDi Challenge used for the segmentation task (10k trajectories each). \textbf{(a)} CONDOR segmentation performance is evaluated by the root mean square error ($\rm RMSE$) of the change point localization as a function of the ground truth value of the change point ($t_i$). The dotted lines represent the overall RMSE for each dimension when using moving windows of width $B_1 = 50$, $B_2 = 40$ and $B_3 = 20$ sequentially (see text). The dash-dotted lines represent the average RMSE for each dimension on all change points identified only with a moving window of width $B_1$ (i.e. for $25 < t_i < 175$). \textbf{(b)} Model classification and \textbf{(c)} $\alpha$-exponent inference with CONDOR on each of the two trajectory segments identified through segmentation. These results are reported by \textbf{(b)} $\rm F_1$ score and \textbf{(c)} $\rm MAE$, respectively, for the first (solid lines) and the second (dashed lines) trajectory segment as a function of $t_i$.}
\label{fig:fig8}
\end{figure}

We tested our algorithm for the segmentation of trajectories on the same datasets used in the AnDi Challenge for the segmentation task \cite{dataset}. For each dimension, a dataset is composed of 10k trajectories 200 time steps long \cite{Andi}. Similar to the datasets for the classification and the inference tasks, these datasets are also corrupted by Gaussian noise corresponding to a finite localization precision. Moreover, in these datasets, at least one parameter between the model and the $\alpha$-exponent changes at a random time $t_i$ within the duration of each trajectory. To predict the change point, we used a moving window of width $B_1 = 50$ at first as explained in Section \ref{sec:NetSegm}. We then used two moving windows of width $B_2 = 40$ and $B_3 = 20$ sequentially to refine the prediction of the change point in trajectories where the previous moving window did not detect a change. The results of our segmentation are reported in Figure~\ref{fig:fig8}. As shown in Figure~\ref{fig:fig8}a, the error in the determination of the change point is strongly influenced by its location in time, reaching its smallest value when the change point is located nearer the center of each trajectory: whilst the overall RMSE for 1D, 2D and 3D trajectories are 53.45, 51.81 and 50.54 respectively, the RMSE reaches average values of 26.00 $\pm$ 0.02 in 1D, 24.20 $\pm$ 0.04 in 2D and 23.01 $\pm$ 0.37 in 3D when the change point is located between 80 and 120 time steps. Interestingly, when the change point location can be determined using larger windows of width $B_1$ (i.e. for $25 < t_i < 175$), even the overall RMSE is significantly reduced (RMSE equal to 39.38 in 1D, 38.41 in 2D and 36.75 in 3D, Figure~\ref{fig:fig8}a), due to CONDOR's better performance in predicting the model and the $\alpha$ exponent when more time points are available. Once the position of the change point is known, CONDOR can be used to classify the model and infer the $\alpha$ exponent for each segment of the trajectory determined by its segmentation. We either used the networks identified in Figure \ref{fig:fig4} for long segments ($T_{\rm max} > 40$) or those trained on short trajectories for short segments ($T_{\rm max} \le 40$). Figure~\ref{fig:fig8}b-c confirms how CONDOR performance in classifying the model and inferring the $\alpha$ exponent increases with the length of the trajectory, which is here determined by the location of the change point in time for each identified trajectory segment. The extra uncertainty introduced by the localization of the change point however leads to slightly lower ${\rm F_1}$ and slightly higher ${\rm MAE}$ values across the dimensions with respect to those in Figure~\ref{fig:fig6}f-h and Figure~\ref{fig:fig7}f-h for trajectories of the same length. \\

\section{Conclusion}
In conclusion, we have introduced a new method (CONDOR) for the characterisation of single anomalous diffusion trajectories in response to the AnDi Challenge \cite{Andi}. Our method combines tools from classical statistics analysis with deep learning techniques. CONDOR can identify the anomalous diffusion model and its exponent with high accuracy, in relatively short times and without the need of any a priori information. Moreover, CONDOR performance is robust to the addition of localization noise. As a consequence, CONDOR was among the top performant methods in the AnDi Challenge (\url{http://www.andi-challenge.org}), performing pregressively better for higher dimension trajectories. At the expenses of an increased computational cost, performance can be further improved by expanding the set of inputs fed to the deep learning algorithm and by employing more advanced network architectures (e.g. recurrent neural networks or convolutional neural networks). While most advanced machine learning techniques often operate as black boxes, we envisage that CONDOR, being partially based on classical statistics analysis, can help characterise the underlying diffusion processes with greater physical insight in many real-life phenomena, from molecular encounters and cellular signalling, to search strategies and the spread of diseases, to trends in financial markets and climate records. 

\section*{Acknowledgements} The authors would like to thank the organisers of the AnDi Challenge for providing a stimulating scientific environment around anomalous diffusion and its characterization. The authors acknowledge sponsorship for this work by the U.S. Office of Naval Research Global (Award No. N62909- 18-1-2170).

\section*{References} 

\bibliographystyle{unsrt}

\end{document}